\documentclass{PoS}

\usepackage{natbib,graphicx}

\def\mnras{MNRAS}
\def\apj{ApJ}
\def\apjl{ApJL}
\def\apjs{ApJS}

\def\prd{PRD}

\title{Weak Lensing with Radio Continuum Surveys}

\ShortTitle{Weak Lensing with Radio Continuum Surveys}

\author{\speaker{Prina Patel}\\
        Department of Physics, University of the Western Cape, Cape Town, 7535, South Africa\\
        E-mail: \email{prina83@gmail.com}}

\abstract{Weak gravitational lensing is a powerful probe of cosmology and has emerged as a key probe for the Dark Universe. Up till now this science has been conducted mainly at optical wavelengths. Current upgraded and future radio facilities will provide greatly improved data that will allow lensing measurements to be made at these longer wavelengths. In this proceedings I show how the larger facilities such as the SKA can produce game changing cosmological measurements even compared to future optical telescopes. I will also discuss how radio surveys can also provide unique ways in which some of the most problematic systematic errors can be  mitigated through the extra information that can be provided in the form of polarisation and rotational velocity measurements. I will also demonstrate the advantages to having overlapping optical and radio weak lensing surveys and how their cross-correlation leads to a cleaner extraction of the cosmological information. Key to the realisation of the great promise of radio weak lensing is the suitable measurements of galaxy shapes in the radio data, either from images or from the visibility data. I shall end with a description of the key issues related to this matter and the radioGREAT challenge which has been proposed to address them. }

\FullConference{EXTRA-RADSUR2015 (*)\\
		20--23 October 2015\\
		Bologna, Italy

                \bigskip
                \hrule
                \bigskip

                \textnormal{(*) This conference has been organized
                  with the support of the Ministry of Foreign Affairs
                  and International Cooperation, Directorate General
                  for the Country Promotion (Bilateral Grant Agreement
                  ZA14GR02 - Mapping the Universe on the Pathway to
                  SKA)}
}

\begin{document}

\section{Introduction}
Weak gravitational lensing is the coherent distortion in the shapes of distant galaxies, and over the past 15 years has emerged as a powerful probe of late time cosmology. Light rays from distant sources are bent by the gravitational potential of objects on the path to an observer, leading to a coherent ellipticity or shear on images of galaxies near each other on the sky. Gravitational lensing is sensitive to the total (i.e. dark and baryonic) matter in the Universe, making it a robust cosmological probe to a large degree without the complications of galaxy formation and bias. With the addition of redshift information weak lensing measurements become sensitive to the geometry of the Universe as well as evolution of structure over cosmic time. Neatly, these latter effects are dependant on the nature of dark energy and/or on the modification the General Relativity on large scales.

Observationally, the field of weak lensing has mainly been concentrated at optical wavelengths, due mainly to much larger number densities achieved by such surveys. Measurements are already maturing at optical frequencies \citep[e.g.][]{2013MNRAS.430.2200K}, and a range of future optical experiments are planned to provide tight constraints on cosmological parameters using this probe (for instance the ground based Large Synoptic Survey Telescope (LSST), \citep{2009arXiv0912.0201L}, and Euclid space telescope, \citep{2012SPIE.8442E..0TL}. See also \citet{LSSTSKASCIENCEBOOK} and \citet{EUCLIDSKASCIENCEBOOK}).

The massive advancement of radio astronomy recently will lead to surveys where the number densities for weak lensing measurements are achieved over the large sky areas required. Specifically, the Square Kilometre Array (SKA) will achieve a number density of $\simeq 5$ galaxies arcmin$^{-2}$ that are well detected and suitably resolved for weak lensing measurements over several thousand square degrees in its initial phase. Initial studies at radio wavelengths have been made by \citet{2004ApJ...617..794C} and \citet{2010MNRAS.401.2572P}, and as shown in \citet{WLSKASCIBK} SKA will be able to provide competitive gravitational lensing measurements. In addition, radio surveys offer unique approaches to weak lensing that are potentially powerful in minimising the main astrophysical weak lensing systematic of intrinsic alignments. 

Weak lensing with data from radio interferometers poses some interesting challenges that originate from the somewhat different way that the data is collected, and then processed to deliver images. Radio interferometers collect the data in the Fourier plane sampled at locations related to the baseline configuration of the interferometer. To realise the promise of weak lensing with these datasets substantial simulation efforts will be required to understand the impact of instrumental systematics as well as the data processing corruptions on galaxy shapes. Shape measurement algorithms in the optical weak lensing community are already very mature, but how they lend themselves to radio data is yet unclear. Indeed, a key question for the radio weak lensing community that requires addressing is whether the shape measurement should be done with the so called visibility (or $uv$) data or with the processed images.   

\section{Prospects of Weak Lensing with Continuum Surveys}
The SKA will provide the ultimate dataset for weak lensing, but as previously mentioned a great amount of algorithm development and exploration of new analysis techniques will be required in the run up to these datasets. A number of SKA pathfinder/precursor instruments lead themselves naturally to these efforts. 

The eMERLIN interferometer in the UK is a key pathfinder telescope for demonstrating the long baseline, high resolution observations suited for weak lensing. Several eMERLIN legacy projects are well suited for pathfinding weak lensing methods and techniques. The e-MERGE\footnote{http://www.e-merlin.ac.uk/legacy/projects/emerge.html} project is a multi-tiered project to observe the GOODS-N field to 0.5 $\mu$Jy rms in the central 100 acrminute$^{2}$ and to 1 $\mu$Jy in the outer 800 arcminute$^{2}$. Complementing e-MERGE is the SuperCLASS\footnote{http://www.e-merlin.ac.uk/legacy/projects/superclass.html} project whose primary science driver is to detect the weak lensing effect around a supercluster of galaxies at redshift $z=0.2$. SuperCLASS will observe a 1.75 deg$^{2}$  region to 4 $\mu$Jy rms. Both these surveys will be conducted at 1.4 GHz with a resolution of 200 mas. 

Continuum surveys conducted on the LOFAR\footnote{http://lofar.strw.leidenuniv.nl/} instrument with the international baselines will be key in testing the weak lensing technique at low frequencies. The CHILES\footnote{http://www.mpia-hd.mpg.de/homes/kreckel/CHILES/index.html} and CHILES con pol\footnote{http://www.aoc.nrao.edu/?chales/chilesconpol/} surveys conducted with the JVLA will be used to to test the novel ideas of using polarisation and radio velocity measurements for radio lensing. Further opportunities exist to explore radio weak lensing with the upgraded JVLA. As pointed out by \citet{2013arXiv1312.5618B} a deep survey conducted with an array configuration that includes long baselines would be extremely useful for pathfinding weak lensing techniques in the radio over larger sky areas approaching 10 deg$^2$.
\subsection{SKA Constraints}
Clearly, the SKA holds the greatest promise for future radio weak measurements. In this section I highlight some forecasts for 2 year SKA continuum surveys as they were initially presented in \citet{WLSKASCIBK}, and refer the interested reader there for further details. 

\begin{figure}
\centering
\includegraphics[width=0.5\linewidth]{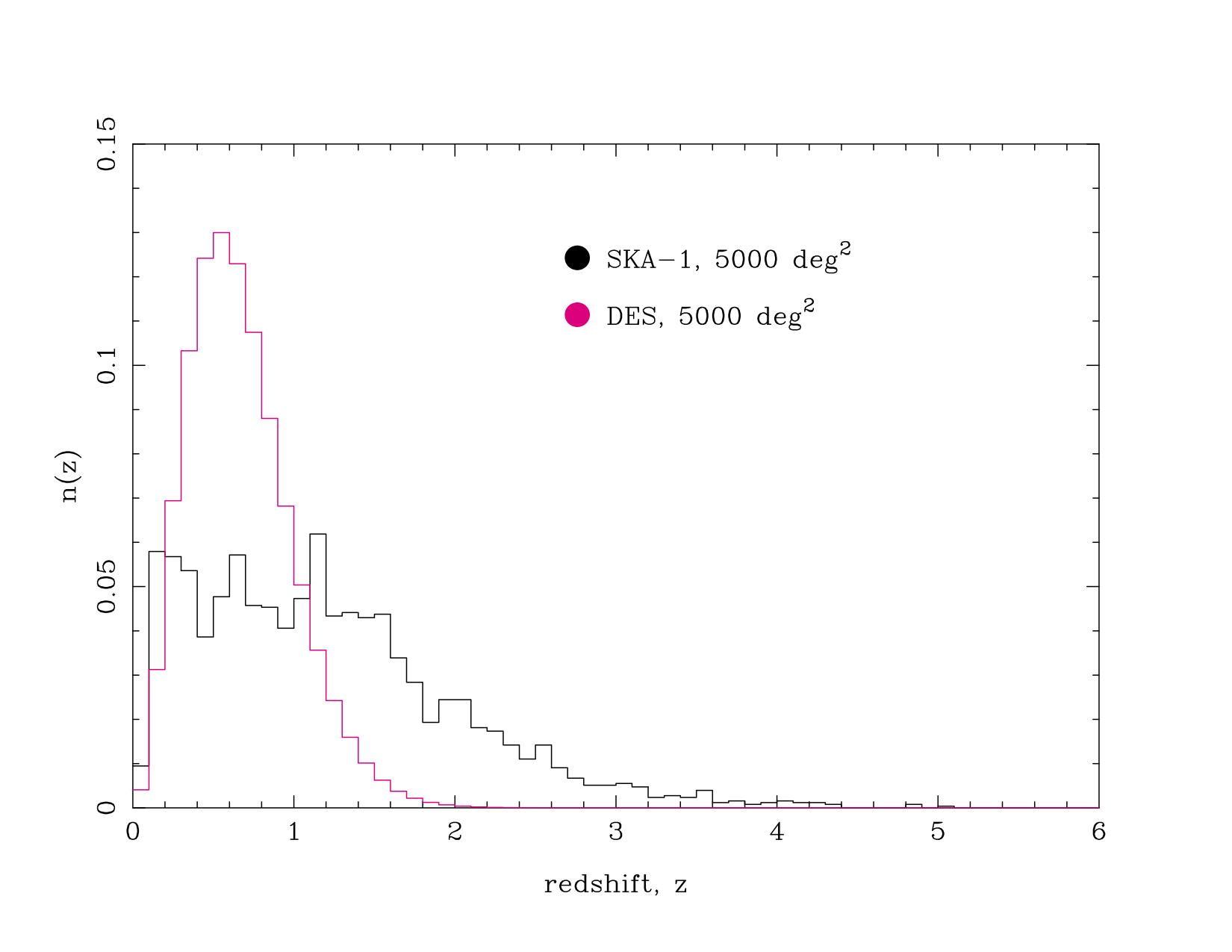}\includegraphics[width=0.5\linewidth]{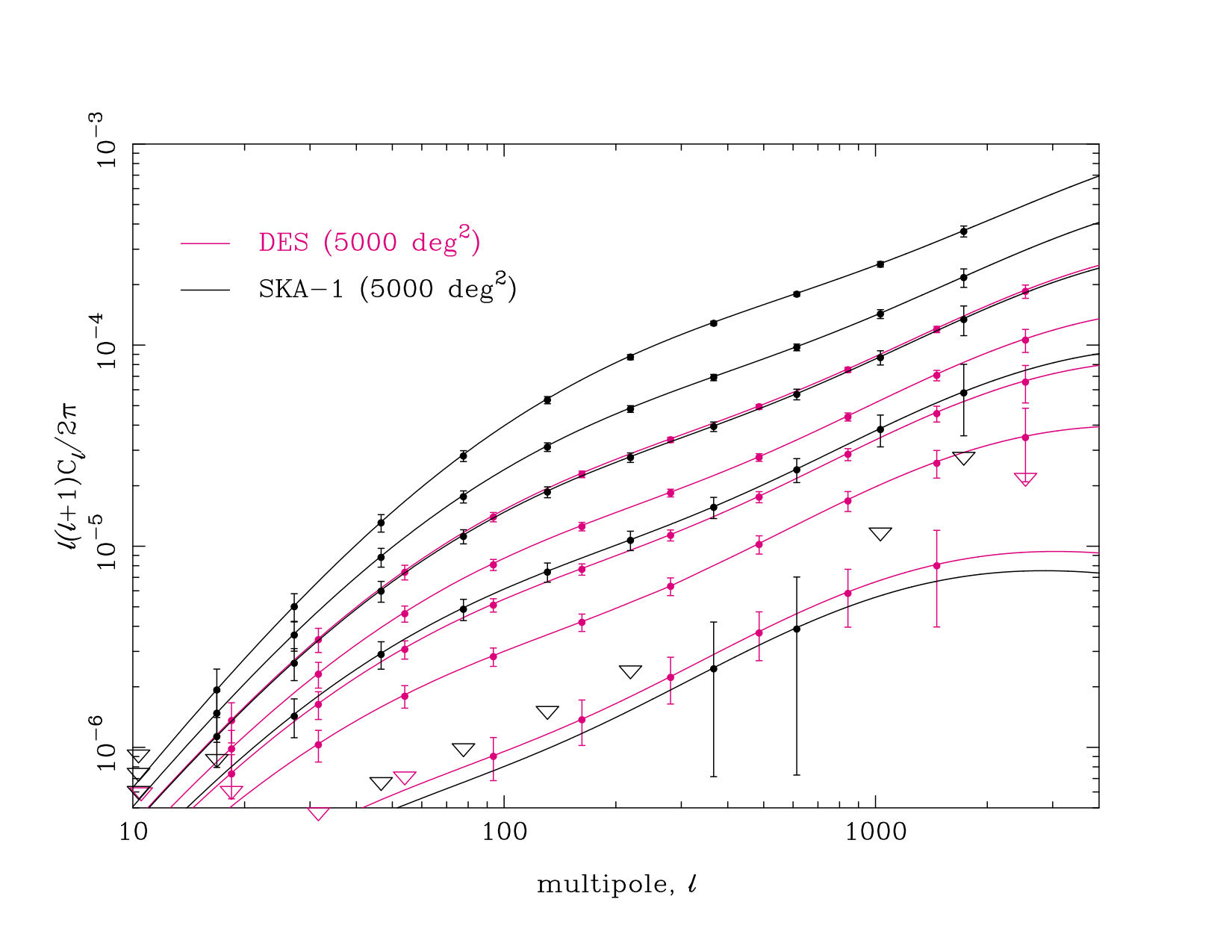}
\caption{\textit{Left panel:} Redshift distribution of source galaxies or a 5000 deg$^{2}$ survey conducted with the SKA1 over 2 years. For comparison the redshift distribution for DES is also shown. \textit{Right panel:} Corresponding constraints for a 5 bin tomographic power spectrum analysis of the two surveys. Image Credit: \citet{WLSKASCIBK}. \label{fig:ska1des}}
\end{figure}

\begin{figure}
\centering
\includegraphics[width=\linewidth]{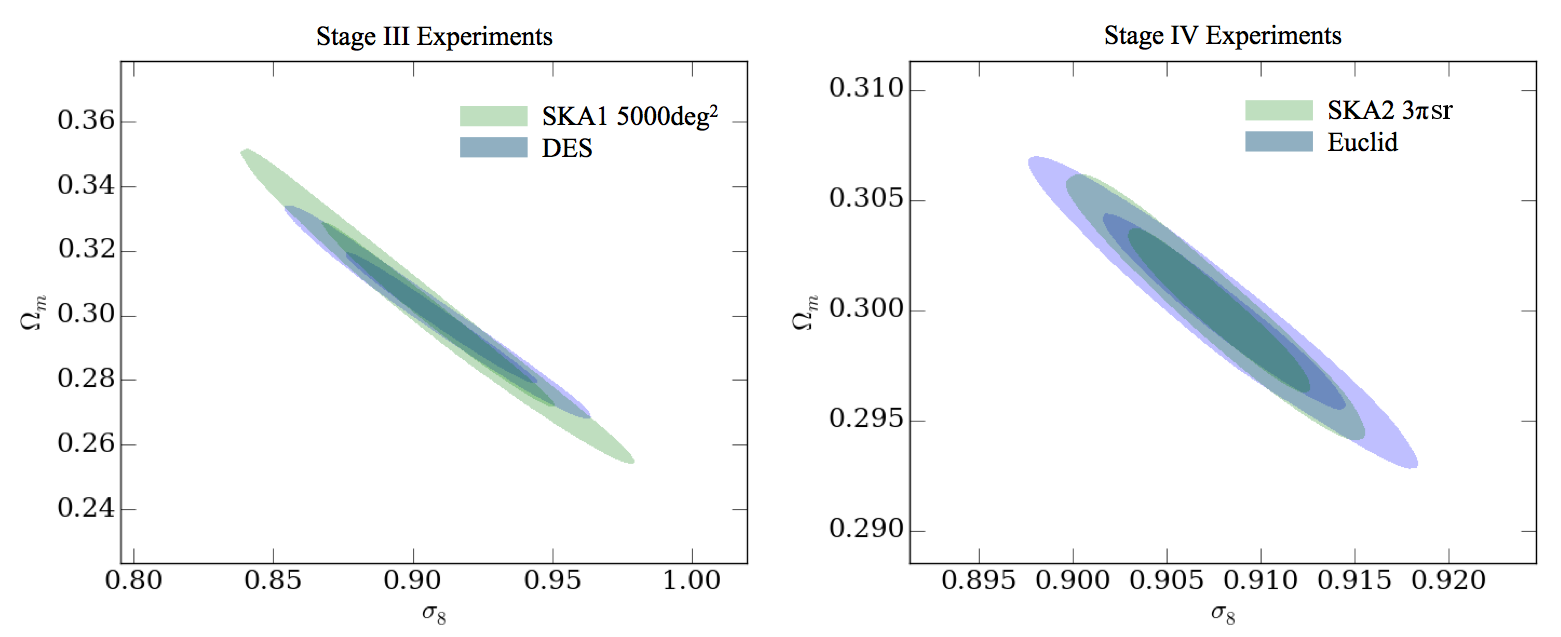}
\caption{Forecasted constraints on $\sigma_{8}-\Omega_{m}$ for 2 year continuum surveys using SKA1 (left) and full SKA (right) performance parameters. For comparison complimentary optical survey constraints are also shown. Image Credit: \citet{2015arXiv150706639H} \label{fig:skacont}}
\end{figure}

Figure \ref{fig:ska1des} illustrates the expected redshift distribution from a 2 year continuum survey conducted with the full SKA1-Mid dish array covering a sky area of 5000 deg$^{2}$ as well as the expected errors on a set of tomographic shear power spectra. These forecasts are generated using the SKA1 performance specifications as outlined in \citet{SKA1IMAGING}. Further, a signal-to-noise threshold of 10 has been applied, a resolution requirement of $\theta_{\textrm{res}}=0.5$ arcseconds imposed and rms dispersion in the intrinsic ellipticity distribution of $\gamma_{\textrm{rms}}=0.3$ is assumed.
For comparison also shown in Figure \ref{fig:ska1des} are the forecasted constraints for the optical Dark Energy Survey\footnote{http://www.darkenergysurvey.org/} (DES) which will be conducted over similar areas and within a similar timescale. The observational parameters used for these forecasts are given in Table \ref{tab:skaparams}. The redshift distribution for the SKA survey extends to high redshift than its optical counterpart, allowing for the power spectra to be measured at higher redshift.
\begin{table}
\caption{Observational parameters used for the power spectrum analysis and cosmological parameter forecasts shown in Figures 1 and 2. Table reproduced from Brown et al. (2015).  \label{tab:skaparams} }
\centering
\begin{tabular}{cccc}
\hline
Survey & $A_{sky}$ (deg$^{2}$) & $n_{gal}$ (arcmin$^{-2}$) & $z_{m}$ \\
\hline
\hline
SKA1-Mid & 5000 & 2.7 & 1.0 \\
DES & 5000 & 6.0 & 0.6 \\
\hline
SKA2 & 30940 & 37 & 1.6 \\
Euclid & 5000 & 30 & 0.9\\
\hline
\end{tabular}
\end{table}
In the left panel of Figure \ref{fig:skacont} we show the corresponding forecasted constraints on the matter density ($\Omega_{m}$) and the matter power spectrum normalisation ($\sigma_{8}$). These constraints are for a 6 parameter $\Lambda$CDM model with no prior, i.e. the constraints come directly from the weak lensing measurements conducted by SKA1-Mid as shown in Figure \ref{fig:ska1des}. Again for comparison the DES constraints are also shown. Details of how these constraints were generated can be found in \citet{WLSKASCIBK}, but note that no systematic errors are included apart from an attempt to model the photometric and spectroscopic redshift errors of the galaxy population. For this particular case, it is assumed that photo-$z$ estimates will be available from overlapping optical surveys with errors of $\sigma_{z}=0.05(1+z)$ out to a limiting redshift of $z=1.5$. For higher redshift galaxies the error is assumed to be $\sigma_{z}=0.3(1+z)$. It is also assumed that spectroscopic redshift information is available for 15\% of sources with $z<0.6$. 
The right hand panel of Figure \ref{fig:skacont} also shows the constraints from the full SKA array with a 2 year, 3$\pi$ steradian survey compared with the optical Euclid survey \citet{2012SPIE.8442E..0TL}. Note these constraints assume that spectroscopic redshifts are available for 50\% of the galaxies with $z<2.0$. 
\subsection{Optical-Radio Cross-Correlations}
Cross-correlating shear estimators from different surveys, e.g. SKA and Euclid/LSST, several systematic errors can be mitigated. For a single survey, the contribution to each shear estimate for each galaxy can be written as
\begin{equation}
\gamma=\gamma+\gamma_{int}+\gamma_{sys},
\end{equation}
where $\gamma$ is the true gravitational shear we seek to measure, $\gamma_{int}$ is the intrinsic ellipticity of the source and $\gamma_{sys}$ is any systematic ellipticity induced by the instrument or data reduction. By correlating shear estimates from a single survey we obtain
\begin{equation}
\langle\gamma\gamma\rangle=\langle\gamma\gamma\rangle+\langle\gamma\gamma_{int}\rangle+\langle\gamma_{int}\gamma_{int}\rangle+\langle\gamma_{sys}\gamma_{sys}\rangle.
\end{equation}
The first term is the gravitational shear signal we are trying to measure, the second term is the GI intrinsic alignment terms \citep{2004PhRvD..70f3526H}. The third term is the II intrinsic alignment term \citep{2000MNRAS.319..649H}, and the final terms accounts for all the contributions from systematics. The final 3 terms are all corruption terms to the true cosmological shear signal. By correlating shear estimates from the radio, $\gamma^{r}$, and optical, $\gamma^{o}$, we obtain
\begin{equation}
\langle\gamma^{r}\gamma^{o}\rangle=\langle\gamma\gamma\rangle+\langle\gamma\gamma_{int}^{r}\rangle+\langle\gamma\gamma_{int}^{o}\rangle+\langle\gamma^{r}_{int}\gamma^{o}_{int}\rangle+\langle\gamma_{sys}^{r}\gamma_{sys}^{o}\rangle.
\end{equation}
We see that the 2 GI terms still survive. The third terms is the correlation between radio and optical shapes which is expected to be less that at a single frequency since the emission mechanisms are different. \citet{2010MNRAS.401.2572P} found this term to be very small at zero lag when comparing matched radio and shear estimates in the HDFN field. More importantly, the final term involving the systematic contributions from the different instruments is expected to be zero since two instruments with different designs and purposes are not expected to have correlated systemics. 

Therefore, by cross-correlating shear estimates from different surveys we are able to obtain a cleaner cosmological signal as we certainly lower the contribution from the intrinsic shape correlations and we remove the systematic contributions which at one frequency can be very large. Since there are many optical surveys that are planned along with various radio surveys with complimentary time scales and survey areas, utilising this cross-correlation technique should further improve the ability with which the cosmological signal can be measured. 
\subsection{Uniqness}
Although cross-correlating with optical surveys is promising for extracting clearer cosmology there are also unique benefits of conducting weak lensing measurements at radio frequencies. In this section I briefly describe how the additional information that comes from polarisation and radial velocity measurements can be used to reduce intrinsic shape biases in the weak lensing analysis. 
\subsubsection{Polarisation}
\citet{2011MNRAS.410.2057B} suggested using the polarisation information of the radio emission of a galaxy as a tracer of the intrinsic position angle. \citet{1992ApJ...390L...5D} have previously shown that the net polarisation position angle is unaffected by lensing, so by utilising the polarisation information we can gain information about the intrinsic alignments of the galaxies. This approach could be used to mitigate the primary astrophysical systematic of intrinsic alignments \citep{2004PhRvD..70f3526H,2000MNRAS.319..649H}, which are a major concern for current and future precision cosmology experiments based on weak lensing measurements. 

In practice this approach is likely to depend crucially on 2 observation parameters: the scatter in the relationship between the polarisation position angle and the intrinsic position angle, denoted $\alpha_{\textrm{rms}}$, and the number of galaxies for which an accurate polarisation position angle can be measured, denoted $n_{\textrm{pol}}$. The quantities are not currently well known and depend further on the polarisation properties of the background galaxies, for example on the mean polarisation fraction $\Pi_{\textrm{pol}}$. A study of local spiral galaxies by \citet{2009ApJ...693.1392S} gives some indication for these parameter values, namely $\alpha_{\textrm{rms}}<15^{\circ}$ and $\Pi_{\textrm{pol}}<20\%$. Clearly, theses parameters need better determination in order for this technique to be feasible. As mentioned above the SuperCLASS, CHILES and CHILES con Pol projects are all testbeds for this method. 
\subsubsection{Radial Velocities}
A second, and similar idea is to utilise rotational velocity measures to provide information regarding the intrinsic shapes of galaxies. First suggested by \citet{2002ApJ...570L..51B} and \citet{2006ApJ...650L..21M}, the idea is to compare the rotation axis of a disk galaxy with the orientation of the major axis. In the absence of lensing these two orientations should be perpendicular, and measuring the departure of perpendicularity is a direct estimate of the shear field at the location of that galaxy. In order to undertake such an analysis would require commensal HI line observations which in principal can be done at no extra cost in terms of telescope time. 

In the limit of perfectly well behaved disk galaxies, the rotational velocity method shares many of the characteristics of the polarisation method in that it is free of shape noise and can therefore be used to remove the intrinsic alignment contamination. In practice, the degree to which the rotation velocity method improves upon standard techniques will depend on observational parameters in analogy to the polarisation method described above. \citet{2013arXiv1311.1489H} have recently proposed an extension to this method using the Tully-Fisher relation to calibrate the rotation velocity shear measurements and thereby reducing the residual shape noise even further. Again existing projects will further investigate the feasibility of this technique. 

\section{radioGREAT: Addressing the Challenges}
In order to realise the potential of weak lensing measurements with current and future radio instruments extremely accurate shape information has to be extracted from real noisy data. In optical experiments the shear measurement process consists of measuring the shapes of the galaxies identified in the images and then combining this shape information into a shear estimate. As a consequence a large number of shape measurement algorithms have been developed and continually tested through the STEP and GREAT programmes \citep[see e.g.][and references therein]{2014ApJS..212....5M}. These shape measurement challenges, along with experience with real data have allowed the methods to be honed and identified areas that require further research for improvement. This has lead to shape measurement in the optical being a mature field and the techniques are now sufficiently advanced that shape measurement induced systematic errors are likely to be sub-dominant to statistical errors in current and near future optical weak lensing surveys. These shape measurement systematics are often parameterised in terms of an additive bias $c$ and a multiplicative bias $m$ on the measured ellipticity $e^{o}$ which is recovered from an input source with known ellipticity $e^{t}$:
\begin{equation}
e^{o}=(1+m)e^{t}+c.
\end{equation}
Using the formalism provided by \citet{2008MNRAS.391..228A} the necessary $m$ and $c$ values can be calculated for upcoming surveys such that the systematic errors are subdominant to statistical ones. 

These existing algorithms, however, have solely been tested on optical and near infrared data and their robustness to issues related with the peculiarities of radio data, e.g. the non-linear deconvolution techniques and sidelobe removal are yet unclear. The two radio weak lensing analyses that have currently been undertaken, both make use of the shapelets \citep{2003MNRAS.338...35R} shape measurement method. \citet{2010MNRAS.401.2572P} make use of the shapelets basis function in the image plane, on images reconstructed using the industry standard CLEAN algorithm. \citet{2004ApJ...617..794C} take advantage of the shapelet basis functions being invariant under Fourier transform, and perform the shear measurement directly in the $uv$ plane. Recent simulated images by \citet{2014MNRAS.444.2893P} for eMERLIN suggest values of $m=0.176$ and $c=0.006$ are achievable with the shapelets in the image domain. For the first phase of the SKA, \citet{2015aska.confE..30P} find that $m=0.28$ and $c=0.001$ is achievable using similar techniques, although the requirements are considerably lower: $m<0.0054$ and $c<0.0073$.

In a similar vain to the aforementioned optical shear testing programmes, radioGREAT\footnote{http://radiogreat.jb.man.ac.uk/} will seek to address the related challenges that shear measurement with radio data presents. In particular the following have been identified as the issues that are crucial:
\begin{itemize}
\item Perhaps the key question that requires addressing is the issue of whether shear measurement should be conducted on deconvolved images or directly in the $uv$ plane with the visibility data. Comparison of image plane and visibility plane shapelets from \citet{2015aska.confE..30P} suggest that visibility plane measurements give improved performance as expected since the imaging step is bypassed altogether. However, the feasibility of working with very large visibility data presents its own challenges.
\item Even outside the scope of weak lensing measurements, there is great effort being made by the radio community to find better alternatives to the CLEAN algorithm, \citep[e.g.][]{2014MNRAS.439.3591C,2014MNRAS.438..768S}. If images are able to be produced to the given fidelity required by weak lensing shear measurement then this naturally allows the infusion of current optical methods. 
\item Since it remains unclear if images of such high fidelity can be created, it is also important that methods be developed to work with the visibility data. As mentioned above, the sheer quantity of data presents challenges that will require suitable algorithms be developed such that the ellipticity estimates of very large numbers of galaxies can be extracted simultaneously from the data in a reasonable timescale. 
\end{itemize} 

The radioGREAT challenge has been accordingly designed to begin addressing the issues mentioned above. Data will be provided in the form of images as well as raw visibilities. The simulated data will use sky models that contain single sources as well as fields. In providing data in these format it is hoped that we can allow the maximum participation from the optical weak lensing community, radio astronomers in the form of imaging experts and also those working in algorithmic development.    
\section{Conclusions}
The field of weak lensing is soon to be open to radio data, and the prospects offered by upcoming experiments offer an exciting new window into the Dark Universe. Ultimately, the SKA will allow for these studies to be conducted over and beyond the traditional optical approach since it will probe galaxy populations at higher redshift where the signal is larger and therefore potentially easier to measure.  Moreover, exploiting the cross-correlation of overlapping optical and radio surveys will allow the extraction of a much cleaner cosmological signal as a number of key systematics can potentially be mitigated. 

Radio continuum surveys will also allow the possibility to measure the weak lensing signal in unique ways that are not available in the optical. Using the polarisation and/or radial velocity measures from HI surveys could in principle allow for the most worrying systematic in weak lensing, the intrinsic alignments, to be mitigated. The feasibility of these techniques are currently being investigated with ongoing experiments. 

To realise the potential of weak lensing at radio wavelengths requires a great deal of development work on analysis techniques and the demonstration of those techniques on real data. A key challenge will be to develop the field of galaxy shape estimation to the level of maturity that is currently enjoyed by the optical weak lensing community.  The radioGREAT challenge has begun in order to draw attention to the highlighted issues and attract researchers from a number of fields to begin addressing them. In conjunction, real data from instruments such as eMERLIN and LOFAR will be available for demonstrating radio weak lensing techniques. 

\section*{Acknowledgements}
PP acknowledges the travel support of the Ministry of Foreign Affairs and International Cooperation, Directorate General for the Country Promotion (Bilateral Grant Agreement ZA14GR02 - Mapping the Universe on the Pathway to SKA)

\bibliographystyle{apj}

\end{document}